\documentclass[superscriptaddress,twocolumn,amsmath,amssymb]{revtex4}
\usepackage{graphicx}
\usepackage{color}
\usepackage{ulem}

\begin{document}

\title{Quantum tunneling with friction}

\author{M. Tokieda}
\affiliation{
Department of Physics, Tohoku University, Sendai 980-8578,  Japan}
\author{K. Hagino}
\affiliation{
Department of Physics, Tohoku University, Sendai 980-8578,  Japan}
\affiliation{Research Center for Electron Photon Science, Tohoku
University, 1-2-1 Mikamine, Sendai 982-0826, Japan}
\affiliation{
National Astronomical Observatory of Japan, 2-21-1 Osawa,
Mitaka, Tokyo 181-8588, Japan}

\begin{abstract}
Using the phenomenological quantum friction models introduced by 
Caldirola-Kanai, Kostin, and Albrecht, we study quantum tunneling 
of a one-dimensional potential in the presence of energy dissipation. 
To this end, we calculate the tunneling probability 
using a time-dependent wave packet method. 
The friction reduces the tunneling probability. 
We show that the three models provide similar penetrabilities to 
each other, among which the Caldirola-Kanai model requires 
the least numerical effort. 
We also discuss the effect of energy dissipation on quantum 
tunneling in terms of barrier distributions. 
\end{abstract}

\maketitle

%%%%%%%%%%%%%%%%%%%%%%%%%%%%%%%%%%%%%%%%%%%%%%%%%%%%%%%%%%%%%%
%%%%%%%%%%%%%%%%%%%%%%%%%%%%%%%%%%%%%%%%%%%%%%%%%%%%%%%%%%%%%%

\section{Introduction}
\label{sec1}
In low-energy heavy-ion fusion reactions, it has been known that 
excitations of the colliding nuclei considerably influence the reaction 
dynamics, that is, fusion cross sections are largely enhanced 
as compared to the prediction of a simple potential 
model \cite{BT98,DHRS98,FUS12,Back14}. 
In order to take into account the excitations during reactions, 
the coupled-channels method has been developed. Many experimental data 
have been successfully accounted for with this method by including a few  
internal degrees of freedom which 
are coupled strongly to the ground state \cite{FUS12,HRK99}. 
However, when a large number of channels are involved, 
the coupled-channels calculations become increasingly 
difficult. 
This is the case, e.g., 
fusion reactions in massive systems, in which many non-collective 
excitations may play an important role \cite{Yusa1,Yusa2,Yusa3,Yusa4}. 

To deal with this problem, many phenomenological models based on the 
classical concept of friction were proposed in 
connection with deep inelastic heavy-ion collisions \cite{FL96}.  
Among them, it has been found 
that the classical Langevin treatment works well 
for fusion reactions and deep inlastic collisions when the 
incident energy is higher than the Coulomb barrier 
\cite{SF,Zagrebaev1,Zagrebaev2}. 
When 
the incident energy is close to the barrier, however, 
the fusion reaction takes place by quantum tunneling. 
Hence, in order to apply these models to low-energy fusion reactions, 
a quantum mechanical extension of the friction models is 
essential. 

Another important issue is to extend the coupled-channels approach 
to massive systems by taking into account the dissipation effects 
and to develop a quantal theory for deep inelastic collision with energy and 
angular momentum dissipations. 
Such theory would be able to describe simultaneously 
dissipative quantum tunneling 
below the Coulomb barrier and deep inelastic collision above the barrier.  
In that way, one may resolve 
a long standing problem of surface diffuseness anomaly in 
heavy-ion fusion reactions, that is, an anomaly that 
a significantly large value for 
the surface diffuseness parameter 
in an inter-nuclear Woods-Saxon potential has to be used in order to 
account for above barrier data of fusion cross 
sections \cite{Newton04,Newton04-2,HRD03}. 
Such theory would also provide a consistent description for deep subbarrier 
hindrance of fusion cross 
sections \cite{Back14}, for which the dynamics after the touching 
of the colliding nuclei play a crucial role 
\cite{Ichikawa15,IM13,Ichikawa07}. 

Quantum friction has attracted lots of attention as a general problem 
of open quantum 
systems \cite{VP11,ZMJP12,GDGM13,IBD14,IBHBD16,KG16,Chou16,EHG16}. 
To date, many attempts at developing a quantum friction 
model have been made. 
They can be mainly categorized into the following two approaches. 
The first is to consider a system with bath, for which 
the environmental bath is often 
simplified as, e.g., a collection of 
harmonic oscillators \cite{UW08,Caldeira81,Caldeira83}. 
The second approach is 
to treat the couplings to the bath implicitly and 
introduce a phenomenological Hamiltonian with which 
the classical equation of motion with a frictional force is reproduced as 
expectation values. 
For this approach, Caldirola and Kanai \cite{CK1,CK2}, Kostin \cite{KO72}, 
and Albrecht \cite{AL75} 
proposed a hermitian Hamiltonian, whose equation of motion contains 
a linear frictional force, while Dekker \cite{DE77} invented an approach with  
a non-hermitian Hamiltonian. 

In this paper, we employ the second approach and investigate quantum tunneling 
in the presence of friction. 
Even though the first approach is more microscopic, 
physical quantities are easier to calculate with the second approach,  
and thus it is easier to gain physical insight into the effect of friction 
on quantum tunneling. 
We particularly consider the three hermitian models for 
quantum friction, that is, 
the Caldirola-Kanai, the Kostin, and the Albrecht models, in order to discuss 
the tunneling problem with quantum friction. 
We mention that 
these Hamiltonians have been applied to a tunneling problem 
\cite{IKG75,Hasse78,CM79,MC84,HH84,BJ92}, but a 
systematic study, including a 
comparison among the models,  
has yet to be carried out with respect to tunneling probabilities. 
In this connection, we notice that Hasse has compared the three models 
for a free wave packet 
propagation and for a damped harmonic oscillator. He 
has shown that the time dependence of the width of a Gaussian wave packet 
varies significantly from one model to another 
while all of these 
three models lead to the same classical equation of motion \cite{HA75}. 
It is therefore not obvious whether the three models lead to 
similar penetrabilities to each other. 

The paper is organized as follows. 
In Sec. \ref{sec2} we briefly introduce the three 
quantum friction models which we employ. 
In Sec. \ref{sec3} 
we present our results for penetrability of a 
one-dimensional barrier. In order to compare among the three models, 
we first carry out 
a detailed study on numerical accuracy of the calculations 
for a free wave packet propagation. 
We then discuss the energy dependence of the penetrability obtained with 
each of these three models. We also discuss the results in terms of barrier 
distribution. 
We finally summarize the paper in Sec. \ref{sec4}.

%%%%%%%%%%%%%%%%%%%%%%%%%%%%%%%%%%%%%%%%%%%%%%%%%%%%%%%%%%%%%%
%%%%%%%%%%%%%%%%%%%%%%%%%%%%%%%%%%%%%%%%%%%%%%%%%%%%%%%%%%%%%%

\section{Quantum Friction models}
\label{sec2}

\subsection{Classical equation of motion}

We consider a system with a particle whose mass is $m$, moving in a 
one-dimensional space $q$ with a potential $V(q)$ 
and a linear frictional force. 
We here consider potential scattering, 
and regard $q$ as the distance between the particle 
and the center of the potential. 
The classical equation of motion for the particle reads
\begin{equation}
\label{eq:classicaleom}
\frac{dp}{dt} + \gamma_0 p + \frac{\partial V}{\partial q}(q) = 0,
\end{equation}
where $p=m\dot{q}$ is the kinetic momentum, the dot denoting the time 
derivative, and $\gamma_0$ 
is a friction coefficient. 
We have assumed that the potential depends only on $q$. 
From the classical equation of motion, Eq. (\ref{eq:classicaleom}), 
the time derivative of the energy $E=p^2/2m + V(q)$ reads
\begin{equation}
\label{eq:classicaldr}
\frac{dE}{dt} =  -\frac{\gamma_0}{m} p^2.
\end{equation}

In constructing the phenomenological quantum friction models, 
Eqs. (\ref{eq:classicaleom}) and (\ref{eq:classicaldr}) have been 
used as a guiding principle, 
that is, it is demanded that the time dependence of the 
expectation values 
obeys the same equations as Eqs. (\ref{eq:classicaleom}) and (\ref{eq:classicaldr}) 
\cite{HA75}. 

\subsection{The Caldirola-Kanai model}

In the Caldirola-Kanai model, the Hamiltonian depends explicitly 
on time as \cite{CK1,CK2}
\begin{equation}
H = \frac{\pi^2}{2m} e^{- \gamma_0 t} + V(q) e^{\gamma_0 t},
\label{eq:H-CK}
\end{equation}
where $\pi$ is a canonical momentum conjugate to $q$. 
The canonical quantization $ [q,\pi] = i\hbar $ with $p=\pi e^{-\gamma_0t}$ 
leads to 
the desired equations,
\begin{equation}
\frac{d}{dt} \left< p \right> + \gamma_0 \left< p \right> + \left< \frac{\partial V}{\partial q}(q) \right> = 0,
\label{eq:ckeos}
\end{equation}
\begin{equation}
\label{eq:ckdisp}
\frac{d}{dt} \left< E \right> = - \frac{\gamma_0}{m} \left< p^2 \right>. 
\end{equation}
Here, 
the expectation value of an operator $O$ is denoted 
as $\left< O \right> = \int dq  \psi^* O \psi$ 
with a wave function $\psi=\psi(q,t)$. 
$ p $ can be regarded as the kinetic momentum 
operator, since the relation $\left< p \right> = m \, (d\left< q \right>/dt)$ 
holds. 
Notice that the kinetic momentum operator depends explicitly on time 
in this model.

Since the Hamiltonian (\ref{eq:H-CK}) is hermitian, 
the probability is conserved with 
the continuity equation of 
\begin{equation}
\frac{\partial \rho}{\partial t} + \frac{ \partial J}{\partial q}e^{-\gamma_0t} = 0,
\end{equation}
where $\rho=|\psi|^2$ and $J=(\hbar/m)\, \Im \left( {\psi^* \partial \psi / \partial q} \right)$ 
are  the probability density and the current, respectively, 
$\Im$ denoting the imaginary part. 

Since the kinetic momentum operator in this model depends 
explicitly on time, the commutation relation 
between the coordinate and the physical momentum 
is of the form
\begin{equation}
\label{eq:uncertain}
[q,p(t)] = i\hbar e^{-\gamma_0t}.
\end{equation}
Hence, the quantum fluctuation disappears as $t \gg 1/\gamma_0$. 
One may consider that this unphysical feature can be neglected 
if one considers only a short time behavior. However, 
the friction is not active in that time regime, since 
the factor $e^{-\gamma_0t}$ determines how 
much the momentum is damped, and thus the dynamics may be rather trivial there. 

\subsection{The Kostin and the Albrecht models} 

In the Kostin and the Albrecht models, 
the momentum operator is kept time-independent, but 
a nonlinear potential $W$ is introduced in the Hamiltonian: 
\begin{equation}
\label{eq:nonlinear}
H = \frac{p^2}{2m} + V(q) + \gamma_0 W. 
\end{equation}
In the Kostin model, the nonlinear potential is taken to be \cite{KO72} 
\begin{eqnarray}
W_{\rm Ko} &=& 
\frac{\hbar}{2i}\left(\ln \frac{\psi}{\psi^*}
-\left\langle 
\frac{\psi}{\psi^*}\right\rangle\right), \\
&=&
\hbar \left[ \Im{\ln{\psi}} - \left< \Im{\ln{\psi}} \right> \right], 
\label{eq:wko}
\end{eqnarray}
while in the Albrecht model it is taken as \cite{AL75}, 
\begin{equation}
W_{\rm Al} = \left< p \right> \left(q - \left< q \right> \right). 
\end{equation}
With the canonical quantization, 
one obtains Eq. (\ref{eq:ckeos}) together with 
\begin{equation}
\frac{\partial \rho}{\partial t} + \frac{\partial J}{\partial q} = 0,
\end{equation}
for both Hamiltonians. 

The energy dissipation for the Kostin model is given by
\begin{equation}
\label{eq:kodisp}
\frac{d}{dt} \left< E \right> = - \frac{\gamma_0}{m} \left< \left( \frac{mJ}{\rho} \right)^2 \right>.
\end{equation}
Kan and Griffin rederived the Kostin Hamiltonian from 
a fluid dynamics point of view \cite{KK76,Ga13}. 
In that context, $mJ/\rho$ in Eq. (\ref{eq:kodisp}) 
is the kinetic momentum, and 
hence Eq. (\ref{eq:kodisp}) is similar to Eq. (\ref{eq:classicaldr}). 
For the Albrecht model, on the other hand, one obtains 
\begin{equation}
\label{eq:aldisp}
\frac{d}{dt} \left< E \right> = - \frac{\gamma_0}{m} \left< p \right>^2,
\end{equation}
as is desired. 
Notice that the energy dissipation is proportional to $\langle p^2 \rangle$ 
in the Caldirola-Kanai and the Kostin models (see Eqs. (\ref{eq:ckdisp}) 
and (\ref{eq:kodisp})), while it is $\langle p \rangle^2$ 
in 
the Albrecht model. 
In the classical limit, these quantities 
are the same to each other, but they may differ in quantum mechanics. 

In Ref. \cite{HA75}, Hasse discussed a generalization of the 
Albrecht model and suggested a better nonlinear potential, $W$,  
which reproduces the classical reduced frequency for 
a damped harmonic oscillator. For simplicity, however, we consider 
only the Albrecht model in this paper. 

\subsection{Generalization for a collision problem} 

In the original models for quantum friction, the friction constant 
$\gamma_0$ is treated to be a constant. 
When considering friction in a collision problem, however, 
we have to introduce a friction form factor $f(q)$, 
since 
the energy dissipation occurs only during interaction. 
That is, the form factor $f(q)$ vanishes outside the range 
of the potential, $V(q)$. 
A naive replacement of $\gamma_0$ in the model Hamiltonians 
with $\gamma_0 f(q)$ does not work 
due to the $q$ dependence in the form factor.
Alternatively, in this paper we consider a time dependent 
friction coefficient $\gamma(t)$ which vanishes after the interaction. 
In the simple form of $\gamma(t)=\gamma_0f(\left< q \right>_t)$,  
the dissipation continuously occurs even after the interaction 
if an incident wave is equally bifurcated into transmitted 
and reflected waves. 
To avoid this undesired behavior, we choose the form,
\begin{equation}
\gamma(t) = \gamma_0 \left< f(q) \right>_t.
\end{equation}
In Ref. \cite{HH84}, Hahn and Hasse discussed a more complex 
form factor but showed that the behavior is quite similar to the simple one.

An extension to the time dependent friction coefficient 
is obvious for the nonlinear potential models; 
just changing $\gamma_0$ to $\gamma(t)$ in Eq. (\ref{eq:nonlinear}).  
For the Caldirola-Kanai model, on the other hand, 
the following modification is necessary 
\cite{S86}: 
\begin{equation}
H = \frac{\pi^2}{2m} \,e^{-\int^t_0 dt' \gamma(t')} + V(q) \,e^{\int^t_0 dt' \gamma(t')}.
\end{equation}
Here we have assumed that the initial time is $t=0$. 
The uncertainty relation is now changed from Eq. (\ref{eq:uncertain}) to
\begin{equation}
\label{eq:uncertain2}
[q,p(t)] = i\hbar \,e^{-\int^t_0 dt' \gamma(t')}.
\end{equation}

Because of these modifications, the three Hamiltonians are now 
nonlinear. It means that the superposition principle is violated. 
We are forced to admit this undesired property, 
since they are inevitable in the present formalism.

%%%%%%%%%%%%%%%%%%%%%%%%%%%%%%%%%%%%%%%%%%%%%%%%%%%%%%%%%%%%%%
%%%%%%%%%%%%%%%%%%%%%%%%%%%%%%%%%%%%%%%%%%%%%%%%%%%%%%%%%%%%%%

\section{Results}
\label{sec3}

To calculate the tunneling probability with 
the three models discussed in the previous section, we integrate 
the time dependent nonlinear Schr\"odinger equation, 
\begin{equation}
i\hbar\frac{\partial}{\partial t}\psi =H\psi.
\label{eq:tdSeq}
\end{equation}
In what follows, we employ the same potential as in 
Ref. \cite{HK04}, that is, 
\begin{equation}
\label{eq:V}
V(q) = V_0 e^{-\frac{q^2}{2s^2}},
\end{equation}
with $V_0 = 100$ MeV and $s = 3$ fm. 
This potential somehow simulates the $^{58}$Ni+$^{58}$Ni reaction, and 
thus we take $mc^2 = 29\times938$ MeV. 

%%%%%%%%%%%%%%%%%%%%%%%%%%%%%%%%%%%%%%%%%%%%%%%%%%%%%%%%%%%%%%

\subsection{Wave packet tunneling without friction}
\label{sec3.1}

Before we introduce the friction, we first discuss the time-dependent 
approach to quantum tunneling. 
For the calculation of the tunneling probability for the time-dependent 
nonlinear Hamiltonians, the usual time-independent approach, which imposes the 
asymptotic plane wave boundary condition, would not be applicable. 
An alternative method is to make 
a wave packet propagate, then observe how 
it bifurcates after it passes the potential region. 

A wave packet is a superposition of 
various waves, each of which has a different energy. 
Hence, 
in order to obtain the tunneling probability for a certain energy,  
one needs either to perform the energy projection \cite{YA97,Diaz-Torres15} 
or to broaden the spatial distribution of the wave packet so that the 
energy distribution becomes narrow \cite{Giraud04}. 
In the former method, 
the tunneling probability is calculated as the ratio of 
the energy distribution of a transmitted wave packet to that of the 
incident one at a fixed energy. 
This method, however, is not applicable in our case, 
since we do not know a priori 
how much energy is lost during a collision at each energy. 
Therefore, we shall employ the latter approach here. 
To this end, it is necessary to know how 
narrow the energy distribution should be in the wave packet in 
order to obtain meaningful results. 

To clarify the effect of finite width in the energy distribution, we take 
the initial wave function in the energy space $\tilde{\psi}_0(E;E_i)$ 
with the Gaussian form,
\begin{equation}
\label{eq:V2}
|\tilde{\psi}_0(E;E_i)|^2 = \frac{1}{\sqrt{2\pi\sigma_E^2}}\,e^{-\frac{(E-E_i)^2}{2\sigma_E^2}},
\end{equation}
where $E_i$ and $\sigma_E$ are the mean energy and the width of the energy 
distribution, respectively. 
Multiplying $\tilde{\psi}_0(E;E_i)$ by $e^{ik(q-q_0)}$ with $E = \hbar^2k^2/(2m)$ 
and making its Fourier transform into the coordinate space, one obtains 
the initial wave function in the coordinate space, $\psi(q,t=0;E_i)$, which 
is consistent with 
the energy distribution given by Eq. (\ref{eq:V2}). 
Such wave function has the mean position of $q_0$. 
Notice that the energy alone does not determine the direction 
of propagation of the wave packet. 
It is determined by the interval of the integration with respect to 
$k$. We consider a propagation of the wave packet from $q_0<0$ towards 
the positive $q$ direction, and thus we 
make the integration from $k=0$ to $\infty$. 
Even though this initial wave function is somewhat different from 
the one used in Refs. \cite{Diaz-Torres15,Giraud04}, we find that this form 
is more convenient in order to discuss a correspondence to the 
time-independent solutions (see Eq. (\ref{eq:exwp}) below).  

By integrating the time dependent Schr\"oringer equation, 
(\ref{eq:tdSeq}), from $t=0$ to $t=t_f$, by which time the bifurcation 
of the wave packet has been completed, 
we calculate the tunneling probability $T_{\rm wp}(E_i)$ as 
\begin{equation}
\label{eq:wptun}
T_{\rm wp}(E_i) = \int^\infty_0 dq\, |\psi(q,t_f;E_i)|^2.
\end{equation}
In implementing the time integration, it is helpful to introduce 
the dimensionless time $\tau\equiv t/t_0$, by measuring the time in units of 
a typical time scale of the problem, $t_0$. 
For this, 
we take $t_0$ as the time taken by a free classical particle 
to travel some distance $L$, that is, $t_0 = L/\sqrt{2E_i/m}$. 
We choose $L$ so that the final mean position of the transmitted 
wave packet at $t=t_0$ is almost independent of $E_i$ (and the friction 
coefficient, $\gamma_0$) for each parameter set.  

When $\sigma_E$ is small enough, it is expected that 
$T_{\rm wp}(E_i)$ is nearly the same as the tunneling probability 
obtained for a certain energy $E_i$, $T_{\rm ex}(E_i)$. 
More generally, the following relation between $T_{\rm wp}$ 
and $T_{\rm ex}$ should hold: 
\begin{equation}
\label{eq:exwp}
T_{\rm wp}(E_i) = \int^\infty_0 dE \,|\tilde{\psi}_0(E;E_i)|^2 T_{\rm ex}(E).
\end{equation}

\begin{figure}[t]
\centering
\includegraphics[clip,width=7cm]{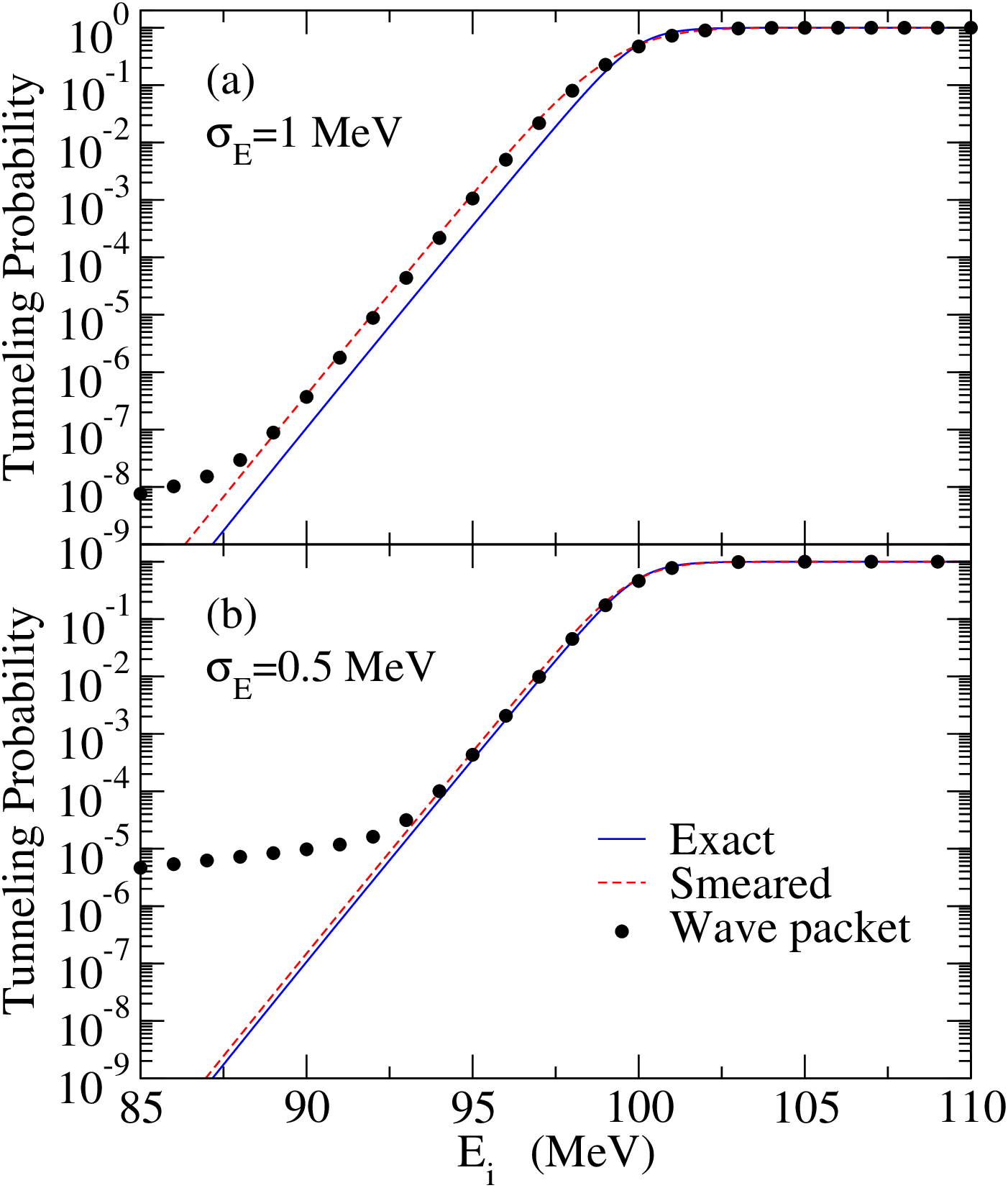}
\caption{Comparison of the tunneling probability of a one-dimensional 
potential obtained with several methods. 
The solid lines show the penetrability obtained with the time-independent 
method, while the filled circles show that with the time-dependent 
wave packet method. 
The dashed lines denote the smeared tunneling probability, 
according to Eq. (\ref{eq:exwp}). 
The upper panel is for the energy width of $\sigma_E = 1$ MeV in the 
wave packet, while the lower panel 
is for $\sigma_E=0.5$ MeV. }
\label{fig:tuntest}
\end{figure}

Without friction, $T_{\rm ex}(E)$ can be calculated with the time-independent 
approach. The upper panel of Fig. \ref{fig:tuntest} shows the 
result for $\sigma_E = 1$ MeV. To solve the time dependent 
Schr\"odinger equation with a wave packet, 
we employ the Crank-Nicholson method together with 
the tridiagonal matrix algorithm \cite{Ko90} with grid sizes of 
$\Delta \tau = 0.00025$ and $\Delta q = 0.01$ fm. 
We take a space of $-100$ fm $<q<100$ fm, and set 
$q_0 = -50$ fm and $L= 165$ fm. 
The solid line shows the penetrability obtained with the time-independent 
method, while the dots are obtained with the wave packet method. 
The dashed line denotes the average penetrability 
according to Eq. (\ref{eq:exwp}). 
One can find that Eq. (\ref{eq:exwp}) is valid until the 
tunneling probability falls below $10^{-7}$. 

In order to improve the agreement between $T_{\rm wp}$ and $T_{\rm ex}$, 
one needs a smaller value of $\sigma_E$. The lower 
panel of Fig. \ref{fig:tuntest} shows the result for $\sigma_E=0.5$ MeV. 
In this case, we enlarge the space to 
$-150$ fm $<q<150$ fm to accommodate a spatially wider wave packet. 
As one can see, the penetrability with the time-dependent wave packet method 
is now in a good agreement with $T_{\rm ex}$ for the tunneling probability 
higher than $10^{-4}$. 
We therefore use $\sigma_E=0.5$ MeV for all the calculations shown below. 

%%%%%%%%%%%%%%%%%%%%%%%%%%%%%%%%%%%%%%%%%%%%%%%%%%%%%%%%%%%%%%

\subsection{Free wave packet evolution with friction}
\label{sec3.2}

\begin{figure}[bt]
\centering
\includegraphics[clip, width=7cm]{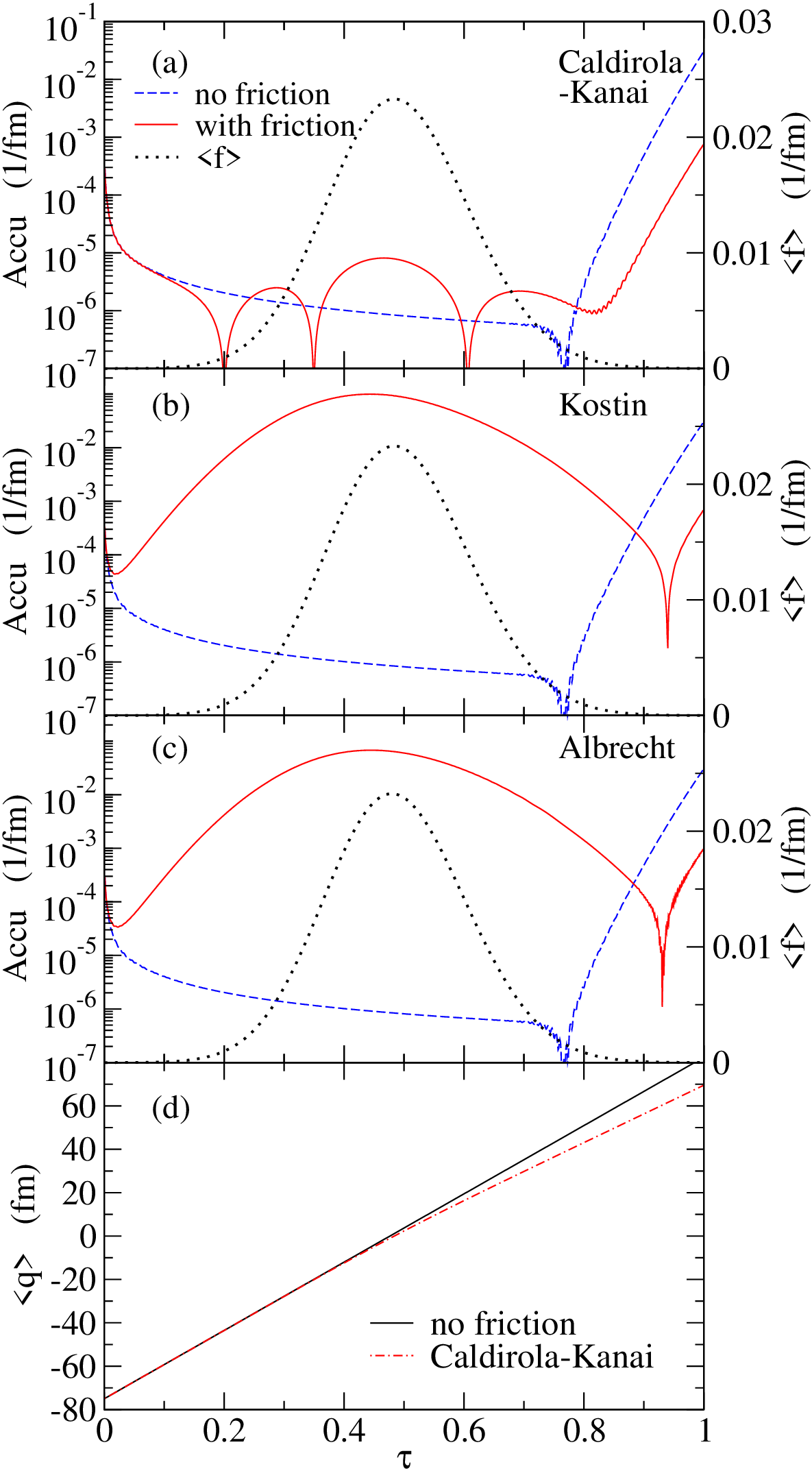}
\caption{Panels (a)-(c): The numerical accuracy 
defined by Eq. (\ref{eq:accu}) 
in the case of the strong friction. 
The dashed and the straight lines 
show the results 
without and with friction, respectively. The dotted lines are 
the expectation value of the form factor, $\langle f(q)\rangle$. 
Panel (d): the time dependence of $\left< q \right>$ for 
the Caldirola-Kanai model. All the calculations are performed 
with $L=160$ fm.}
\label{fig:NAs}
\end{figure}

\begin{figure}[bt]
\centering
\includegraphics[clip, width=7cm]{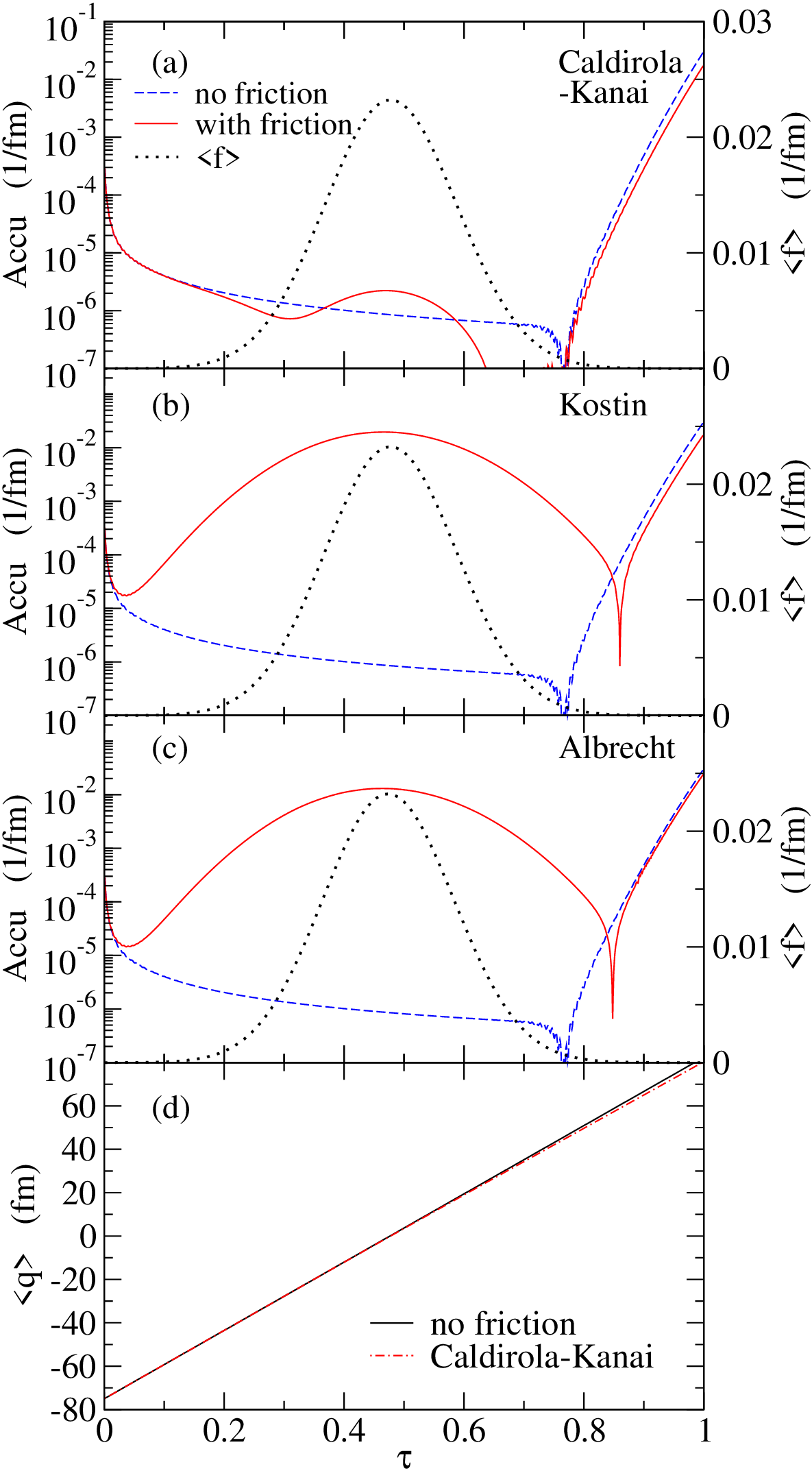}
\caption{Same as Fig. {\ref{fig:NAs}}, but in the case of the weak friction.}
\label{fig:NAw}
\end{figure}

In order to discuss the value of 
a friction coefficient as well as numerical accuracy of the 
calculations, we next 
consider free wave packet in 
this subsection. As discussed in Sec. \ref{sec2}, the potential and the 
corresponding friction form factor should have a similar 
range. In this paper we simply employ the same form 
for the form factor 
as that for the potential, Eq. (\ref{eq:V}),
\begin{equation}
\label{eq:formfactor}
f(q) = \frac{1}{\sqrt{2 \pi s^2}}\,e^{-\frac{q^2}{2s^2}}.
\end{equation}
Here $f(q)$ is normalized so that $\gamma_0$ is interpreted as the strength 
of friction. Note that the dimension of $\gamma_0$ is altered 
from inverse time to velocity.

The friction strength $\gamma_0$ is determined based on the 
amount of energy loss. Since the energy loss depends on energy, 
we choose the barrier top energy, $E_i=100$ MeV, as a reference. 
With the mean energy of the transmitted wave, $E_f$, 
the energy loss $E_{\rm loss}$ is given by $E_{\rm loss} = E_i-E_f$.
We here consider weak and strong friction cases, for 
which $E_{\rm loss}$ is $5$ MeV and $30$ MeV, respectively. 
These are realized when $\gamma_0$ is chosen as listed in 
Table \ref{table:friction}. 

\begin{table}[h]
\centering
\begin{tabular}{c|ccc}
\hline
\hline
strength & Caldirola -Kanai & Kostin & Albrecht \\
\hline 
weak $(E_{\rm loss}$ = 5 MeV) & 2.14 & 2.16 & 2.17 \\
\hline
strong ($E_{\rm loss}$ = 30 MeV) & 13.8 & 14.0 & 14.0 \\
\hline
\hline
\end{tabular}
\caption{The dimensionless friction coefficient $\gamma_0/c$, $c$ being 
the speed of light, for the weak and strong friction cases. All the values 
are given in units of 10$^{-3}$.}
\label{table:friction}
\end{table}

Using these friction coefficients, the accuracy of integration of the 
time dependent nonlinear Schr\"odinger equation is tested by checking 
how well the equation of motion, Eq. (\ref{eq:ckeos}), 
is reproduced. 
The equation cannot be solved in the same way as in Sec. \ref{sec3.1}, 
since the matrix is no longer in a tridiagonal form due to the nonlinearity.
Instead, we carry out the numerical integration in the following way. 

The discretized Schr\"odinger equation may be given by
\begin{equation}
\label{eq:cn}
i\hbar \frac{\psi^{n+1}-\psi^{n}}{\Delta t} = \frac{H^{n+1}\psi^{n+1}+H^{n}\psi^{n}}{2},
\end{equation}
at the $n$-th time grid. Here $H$ is the Hamiltonian which depends on $\psi$. 
In our calculation, we simply neglect the time dependence of 
the Hamiltonian and obtain,
\begin{equation}
\label{eq:SA}
i\hbar \frac{\psi^{n+1}-\psi^{n}}{\Delta t} = H^{n}\frac{\psi^{n+1}+\psi^{n}}{2}.
\end{equation}
We integrate this equation with grid sizes of 
$\Delta \tau = 0.00025$, $\Delta q = 0.01$ fm for the Caldirola-Kanai 
and the Kostin models, and $\Delta \tau = 0.00015$, $\Delta q = 0.01$ 
fm for the Albrecht model.

A care must be taken in integrating the equation for the Kostin model. 
The nonlinear potential Eq. (\ref{eq:wko}) is nothing but the 
phase of a wave function, and hence a naive estimation leads to 
discontinuities in the potential. However, one 
can estimate the continuous phase by the following 
definition \cite{Ga13} (see also Ref. \cite{KG16}):
\begin{equation}
\label{eq:log}
\arg \psi(q,t) = \Im \ln{\psi(q,t)} + 2\pi (n_{+}-n_{-}),
\end{equation}
where $n_{+}$ and $n_{-}$ are the number of crossing the discontinuous 
points from $\pi$ to $-\pi$ and from $-\pi$ to $\pi$, 
respectively. We take $q=0$ as a reference and compute $n_{+} \left( n_{-} \right) $ by counting the point where adjacent phase differs 
by less than $-4$ $ \left( {\rm more \ than} \ 4 \right) $.

Our numerical test is carried out for $E_i = 100$ MeV. 
We compare the following quantity with that of the no friction: 
\begin{equation}
\label{eq:accu}
Accu \equiv \frac{1}{\hbar}\, \left|\frac{d}{d\tau} \left< p \right> 
+ \gamma t_0 \left< p \right>\right|.
\end{equation}
Since we do not consider the potential in this subsection, this quantity 
vanishes if the equation of motion is fully satisfied. 
The accuracy for the strong friction case is shown in Figs. 
\ref{fig:NAs} (a)-(c) for the three friction models. 
The expectation value of the form factor, $\left< f \right>$, is also 
shown to illustrate the effect of nonlinearity on numerical accuracy. 
The corresponding $\left< q \right>$ as a function of $\tau$ 
is also shown in Fig. \ref{fig:NAs} (d) 
for the Caldirola-Kanai model (the results for the other two 
models are almost the same and are not shown 
in the figure). 
We have verified that the probability is conserved within a numerical 
accuracy for all the calculations. 
It is found that the Caldirola-Kanai model can be integrated as 
accurately as the no friction case. 
In contrast, the reproduction of the equation of motion is less 
satisfactorily with the Kostin and the Albrecht models due to the nonlinearity 
of the equations. 
This is expected if the nonlinearity due to the form factor 
plays a much less important role 
as compared to the nonlinearity of the equation itself, 
since 
the nonlinearity of the Caldirola-Kanai 
model is caused only by the friction form factor, Eq. (\ref{eq:formfactor}). 
Actually, we have verified that the accuracy remains almost the same as 
Fig. \ref{fig:NAs} even without 
the form factor for all the models. 

Notice that the increase of $Accu$ at large $\tau$ is due to 
the finiteness of our space, that is, $q$ is limited in the range 
of $-$150 fm $\leq q \leq$ 150 fm. 
 An accumulation of numerical errors is rather small, 
as no increase is observed when $L$ is taken to  be small enough 
so that the tail of the wave packet does not reach  
the edge of the box at $\tau=1$ while keeping the number of step 
in the $\tau$ integration to be the same. 

The accuracy of the nonlinear potential models does not improve even for 
the weak friction, as shown in Figs. {\ref{fig:NAw}} (a)-(c). 
Even though the absolute value of $Accu$ is slightly reduced in this 
case, the $\tau$-dependence remains almost the same. 

We should note that the damping from a bound excited state to the 
ground state can successfully 
be described with the Kostin model \cite{Ga13,Ch15}. 
Actually we also have verified it for a harmonic oscillator 
with the same grid sizes.  
An application to scattering problems seems more 
difficult with our present numerical method.

%%%%%%%%%%%%%%%%%%%%%%%%%%%%%%%%%%%%%%%%%%%%%%%%%%%%%%%%%%%%%%
%%%%%%%%%%%%%%%%%%%%%%%%%%%%%%%%%%%%%%%%%%%%%%%%%%%%%%%%%%%%%%

\subsection{Quantum tunneling with friction}
\label{sec4}

\begin{figure}[tb]
\centering
\includegraphics[clip, width=7cm]{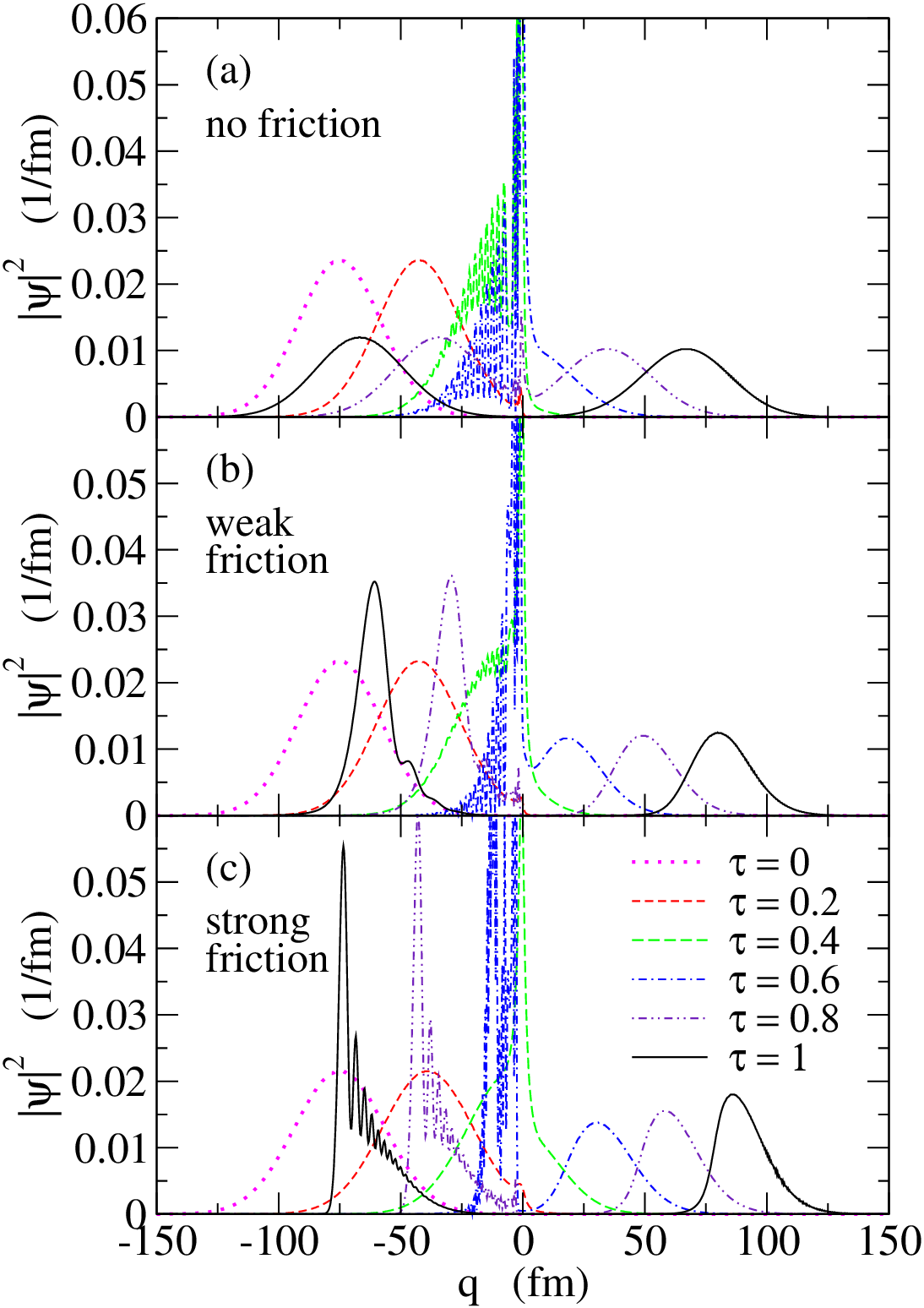}
\caption{
The time-evolution of a wave packet for a tunneling problem in the 
Caldirola-Kanai model. 
The upper, the middle, and the lower panels correspond to the case 
of without friction, the weak friction, and the strong friction, 
respectively. 
The initial energy $E_i$ and $L$ are set to be 
$E_i = 100$ MeV and $L=165$ fm for the no friction, 
$E_i = 103$ MeV and $L=165$ fm for the weak friction, 
and $E_i = 120$ MeV and $L = 185$ fm for the strong friction.}
\label{fig:wptun}
\end{figure}

\begin{figure}[tb]
\centering
\includegraphics[clip, width=6.5cm]{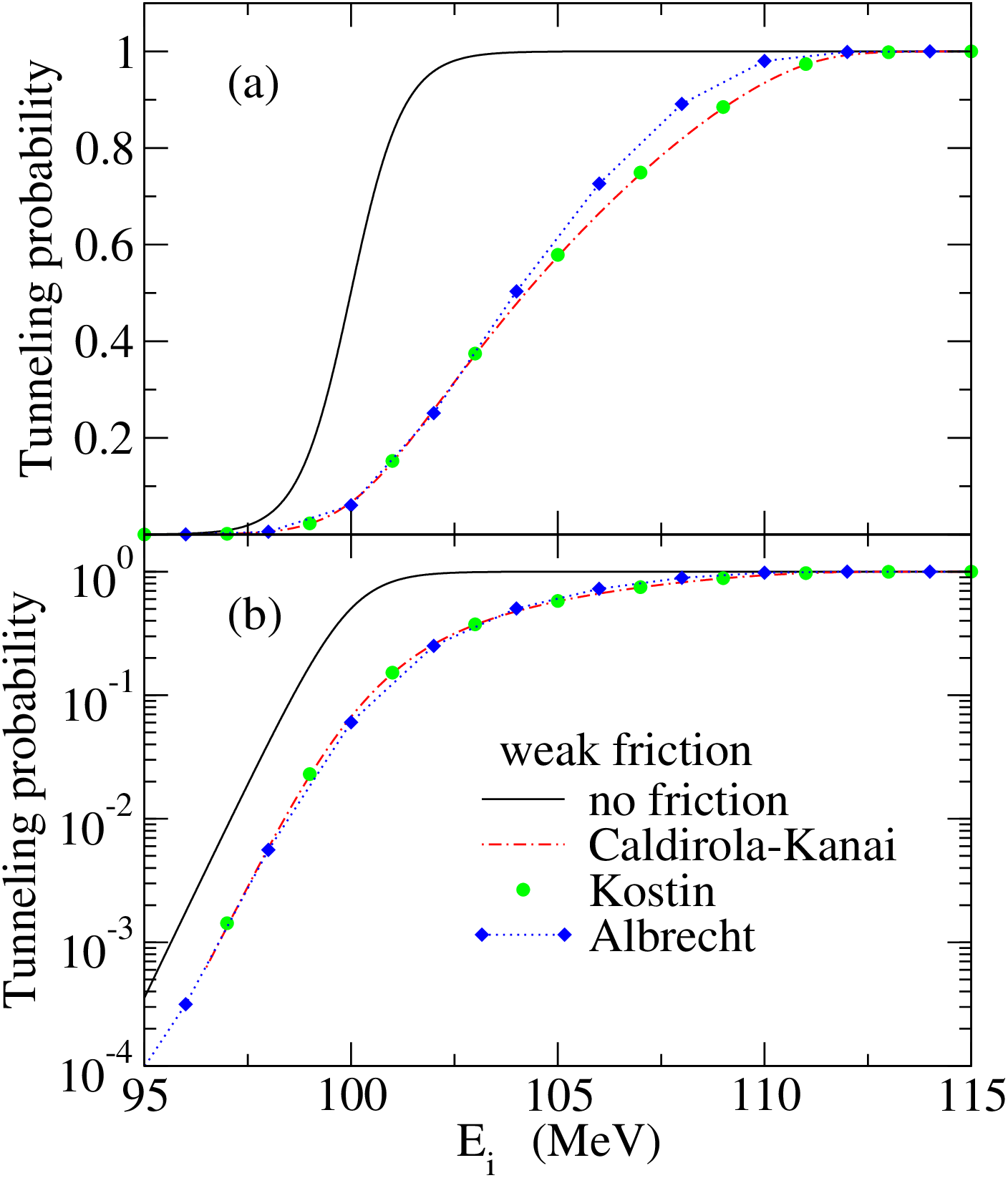}
\caption{The energy dependence of the tunneling probability for 
the Caldirola-Kanai (the dot-dashed lines), the Kostin 
(the filled circles), and the Albrecht 
(the dotted lines with filled diamonds) models with the weak friction. 
The upper panel is in the linear scale, while the lower panel 
is in the logarithmic scale. The result without friction is also shown 
by the solid lines for a comparison.}
\label{fig:Wtun}
\end{figure}

\begin{figure}[tb]
\centering
\includegraphics[clip, width=6.5cm]{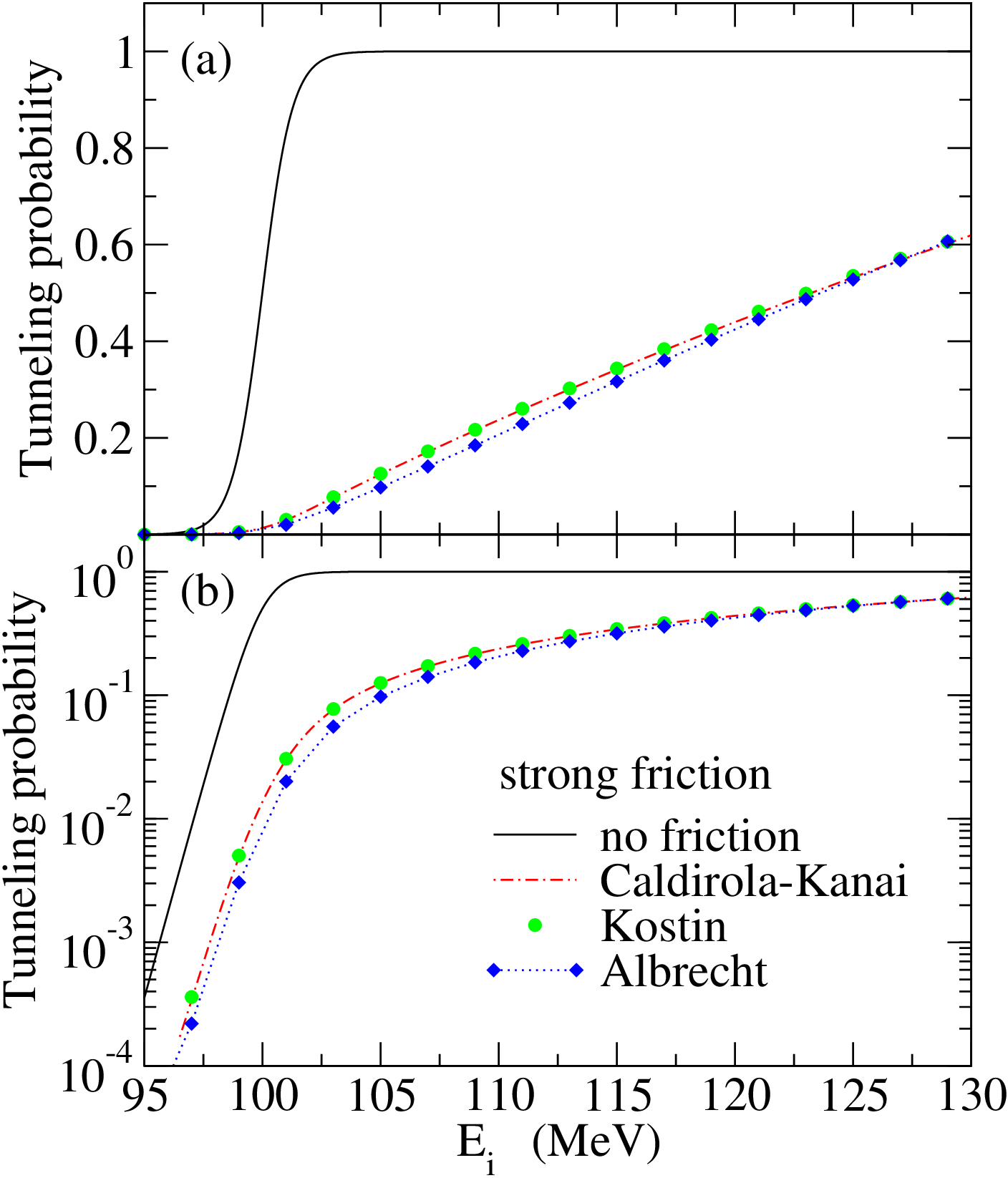}
\caption{Same as Fig. {\ref{fig:Wtun}}, 
but with the strong friction.}
\label{fig:Stun}
\end{figure}

Let us now discuss 
dissipative quantum tunneling, i.e., 
quantum tunneling in the presence of friction. 
For an illustration, Fig. \ref{fig:wptun} shows 
the time-evolution of the wave packet 
for the Caldirola-Kanai model. 
The initial mean energy is chosen so 
that the transmitted wave packet has an appreciable amount. 
The calculations are performed 
with $E_i = 100$ MeV and $L=165$ fm for the no friction, 
$E_i = 103$ MeV and $L=165$ fm for the weak friction, 
and $E_i = 120$ MeV and $L = 185$ fm for the strong friction.
The behavior is quite similar also for the nonlinear potential models. 
Notice that the reflected and the transmitted wave packets 
at $\tau$ = 1 largely deviate from a symmetric Gaussian shape 
in the presence of friction. 

Figs. \ref{fig:Wtun} and \ref{fig:Stun} 
compare the tunneling probability as a function of energy obtained 
with the three friction models for the weak and the strong 
friction cases, respectively. 
We plot only the tunneling probability larger than $10^{-4}$,  
according to the discussion in Sec. \ref{sec3.1}.
One can see that 
the tunneling probability of the three models is nearly the same, even though 
there might be a possibility that 
the results of the Kostin and the Albrecht models suffer from 
numerical errors with the present setup of numerical calculations 
(see the discussion in Sec. \ref{sec3.2}). 
It is interesting to notice that 
the Caldirola-Kanai and the Kostin models lead to almost the same results 
to each other, while the result of the Albrecht model 
slightly deviates from the other two models. 
Even though the exact cause of this different behavior is not known, 
a possible origin may be the fact that 
the energy dissipation is slightly different between the Albrecht model 
and the Caldirola-Kanai/Kostin models 
(see the discussion 
below Eq. (\ref{eq:aldisp})). 

\begin{figure}[tb]
\centering
\includegraphics[clip, width=6.5cm]{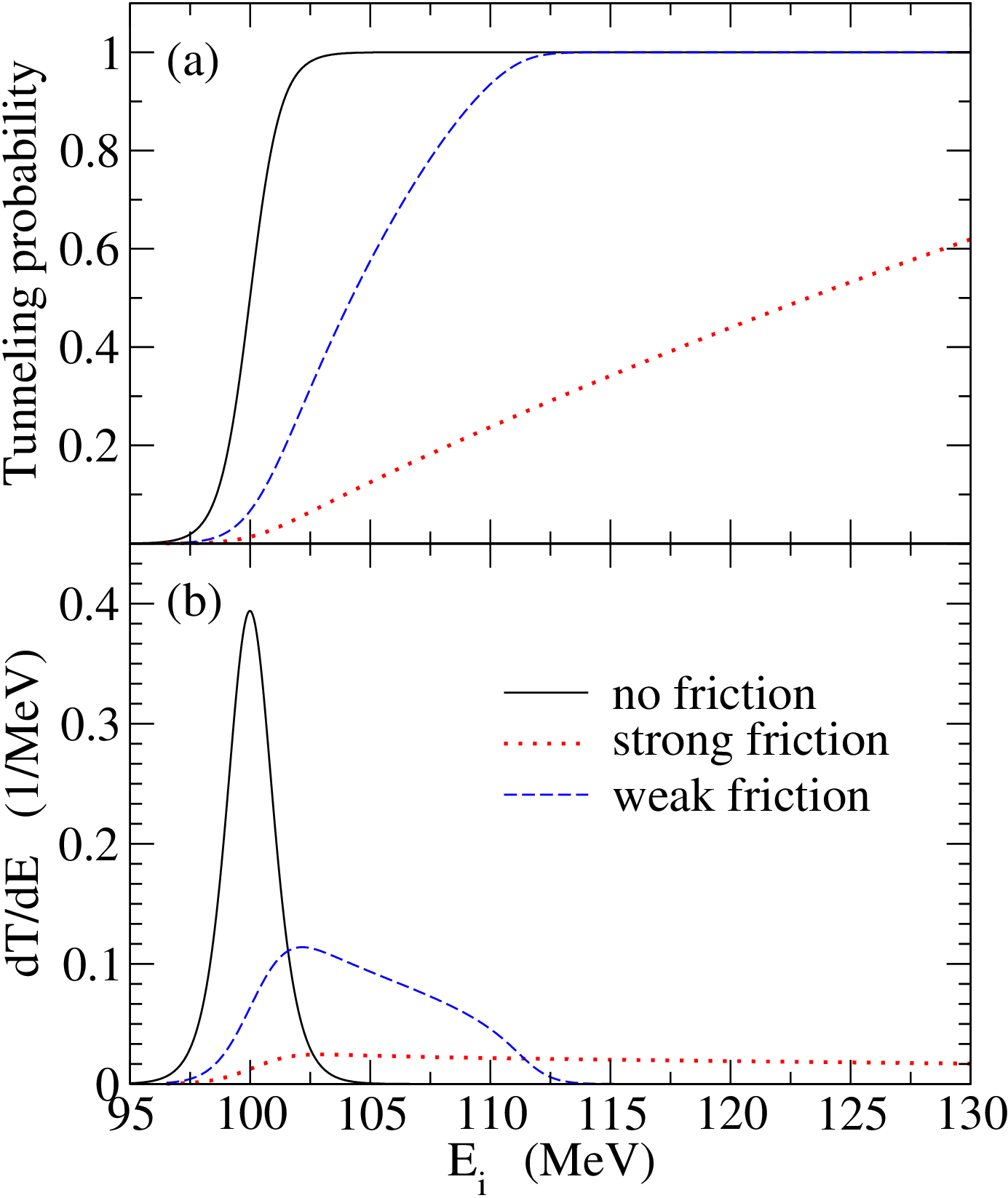}
\caption{
Upper panel: the penetrabilities for the 
Caldirola-Kanai model as a function of incident energy. 
The solid line corresponds to the no friction case. 
The dashed and the dotted lines are for 
the weak and strong friction cases, respectively. 
Lower panel: the corresponding barrier distribution defined 
as the first energy derivative of the penetrability. 
}
\label{fig:BD}
\end{figure}

In what follows, we focus only on the result of the 
Caldirola-Kanai model. 
As can be seen in the upper panel of 
Fig. \ref{fig:BD}, 
the stronger the friction is, the lower the tunneling probability results in. 
This behavior is consistent with the results of Ref. \cite{IKG75} 
for a rectangular barrier unless the tunneling probability is 
extremely small. Ref. \cite{IKG75} showed that the tunneling probability 
is not affected by friction at energies well below the barrier. 
Whereas the numerical accuracy has yet to be estimated in order to draw 
a conclusive conclusion concerning the role of friction in quantum tunneling 
at deep subbarrier energies, 
we simply could not confirm the result of 
Ref. \cite{IKG75} because a finite width in the wave packet prevents us 
to go into the deep subbarrier energy region (see the lower panel of 
Fig. \ref{fig:tuntest}). 
In Ref. \cite{MC84}, McCoy and Carbonell argued that the 
tunneling probability is either increased or decreased by fiction 
depending on the magnitude of the barrier height and width. 
We do not confirm their results either, partly because we 
do not include the fluctuation term in the Hamiltonian. 

In order to gain a deeper insight 
into the role of 
friction in quantum tunneling, we next discuss a barrier distribution. 
In the field of heavy-ion subbarrier fusion reactions, 
the so called fusion barrier distribution has been widely 
used in analyses of experimental data \cite{DHRS98,Leigh95}. 
This quantity is defined as the 
second energy derivative of the product of the incident energy $E$ 
and fusion cross sections $\sigma_{\rm fus}$, that is, 
$d^2(E\sigma_{\rm fus})/dE^2$ \cite{RSS91}, and has provided a 
convenient representation in order to study the underlying dynamics 
of subbarrier fusion reactions. 
For the tunneling problem, this quantity corresponds to the 
first energy derivative of the penetrability, $dT/dE$ \cite{FUS12}. 
The barrier distribution for the Caldirola-Kanai model is shown in 
the lower panel of Fig. \ref{fig:BD}. 
Whereas the barrier distribution shows a symmetric peak 
in the case of no friction, some strength is shifted towards higher 
energies as the strength of the friction increases and the 
barrier distribution becomes structured. 
It is interesting to notice that a similar behavior has 
been obtained in coupled-channels 
calculations for fusion in relatively heavy systems, 
such as $^{100}$Mo+$^{100}$Mo \cite{RGH06}. 

The barrier distribution indicates that 
the energy damping during tunneling results in a increased 
effective barrier, whose height is thus energy dependent 
and is 
determined by the 
strength of friction. 
This leads us to two different points of view for dissipative 
quantum tunneling. From one view point, the incident energy 
is damped by friction while a wave packet traverses 
towards a fixed barrier. This can be interpreted in a different way 
as that the effective barrier increases dynamically due to the friction 
for a fixed value of incident energy. The barrier distribution shown in 
Fig. \ref{fig:BD} well represents this dynamical point of view of 
friction.

%%%%%%%%%%%%%%%%%%%%%%%%%%%%%%%%%%%%%%%%%%%%%%%%%%%%%%%%%%%%%%
%%%%%%%%%%%%%%%%%%%%%%%%%%%%%%%%%%%%%%%%%%%%%%%%%%%%%%%%%%%%%%

\section{Summary}
\label{sec5}

We have investigated 
the effects of friction on quantum tunneling by applying 
the three friction models, the Caldirola-Kanai, the Kostin, and 
the Albrecht models, to a one-dimensional tunneling problem. 
We have studied the 
energy dependence of the tunneling probability 
obtained as the barrier penetration rate of a wave packet, 
whose initial energy variance is set to be small enough.
In order to limit a region where the dissipation is active, 
we have introduced the time dependent friction coefficient. 
We have shown 
that the friction tends to prevent the wave packet from penetrating 
the barrier, and thus the penetrability decreases as a function 
of the strength of friction. 
We have found that the three models lead to similar penetrabilities 
to each other. 
We have also discussed the effect of friction on quantum tunneling 
in terms of barrier distribution and have shown that the 
barrier distribution becomes structured due to friction by 
shifting effective barriers towards higher energies. 
Among the three models which we considered in this paper, we have 
found that the numerical accuracy can be most easily handled 
with the Caldirola-Kanai model. 

Very recently, it has been found experimentally that 
heavy-ion multi-nucleon transfer 
processes in $^{16,18}$O, $^{19}$F + $^{208}$Pb reactions populate highly 
excited states in the target-like nuclei \cite{Rafferty16}. 
One may be able to describe such processes by extending 
the friction models considered in this paper to 
multi-channel cases. We are now working towards this direction, and we 
will report our results in a separate paper. 
Another interesting future direction is to include the random force 
term to the quantum friction Hamiltonians and investigate its effect 
on quantum tunneling. For this purpose, a proper quantization of 
the fluctuation term will be needed. 

%%%%%%%%%%%%%%%%%%%%%%%%%%%%%%%%%%%%%%%%%%%%%%%%%%%%%%%%%%%%%%
%%%%%%%%%%%%%%%%%%%%%%%%%%%%%%%%%%%%%%%%%%%%%%%%%%%%%%%%%%%%%%

\end{document}